\documentstyle[12pt]{article}
\begin{document}
\def\th{\theta}
\def\eps{\epsilon}
\def\vk{{\bf k}}
\def\de{\Delta}
\def\par{\partial}
\def\cj{{\Im}}
\def\kb{k_B\beta}
\def\kbtc{k_BT_c}
\def\angs{\stackrel{\rm o}{\rm A}}
\def\beq{\begin{equation}}
\def\enq{\end{equation}}
\def\beqn{\begin{eqnarray}}
\def\eenq{\end{eqnarray}}
\def\pl{\parallel}
\baselineskip22pt
\begin{center}
{\large{\bf Impurity Substitution in Bismuth and Thallium Cuprates: 
            Suppression of T$_c$ and Estimation of Pseudogap}}\\  
\vspace{0.8cm} 
{\bf Biplab Chattopadhyay}$^{\star}$, {\bf B. Bandyopadhyay}, 
{\bf Asok Poddar}, {\bf P. Mandal}, \\ {\bf A. N. Das} and {\bf B. Ghosh}\\
\vspace{0.2cm} 
{Saha Institute of Nuclear Physics,
     1/AF Bidhannagar, Calcutta - 700 064, INDIA}  
\end{center} 
\vspace{1cm}

\begin{abstract} 

Suppression of $T_c$ in bilayer bismuth and thallium cuprates, by
substitution of Co impurities at Cu sites, are taken for examination.
$T_c$ suppression data on differently doped Bi2212 and Tl2212 are
analysed within the unitary pair-breaking formalism due to Abrikosov
and Gorkov, by fitting data points to a phenomenological relation valid
for weak coupling $d$-wave superconductors. Values of the pseudogap
magnitude at each doping are thereby estimated within a ``fermi-level 
density of states suppression" picture. Pseudogap magnitude
from our estimation is observed to have a correspondence with 
a related characteristic temperature $T^\star$ obtained by
thermoelectric power measurements. Effects of pseudogap, on the
density of states, is studied by calculating the susceptibility 
which shows a broad peak at high temperature. This peak feature
in susceptibility is indicative of an unusual metallic state which 
could further be explored by systematic other measurements. 

\end{abstract}
 
\vspace{10pt} 
      
\vspace{12pt} 

\indent~~~~~~~~~~ formalism, pseudogap. 
\vfill

\noindent \rule{15.73cm}{0.1mm}\\
{\small $^\star$email: biplab@cmp.saha.ernet.in} 

\newpage 
\centerline{\large{\bf I.  Introduction}} 

In high-$T_c$ layered superconductors (HTLS), studying the 
suppression of $T_c$ by impurity substitution at the Cu-site 
of the CuO$_2$ plane is of considerable importance, because 
it can provide useful information regarding the symmetry of 
the superconducting order parameter, and could also throw light 
on the recent striking issue of the observation of a normal state 
pseudogap in underdoped materials. In fact, in the recent years, 
there have been growing interest in the studies of suppression of 
$T_c$ in HTLS by intentional substitution of both magnetic 
(Co,~Ni) as well as non-magnetic (Zn) impurities at copper (Cu) 
sites. These impurities essentially cause pair breaking 
\cite{ishida,tallon} leading to the reduction of $T_c$ while 
leaving the carrier concentration in the CuO$_2$ plane 
unaltered \cite{maeda,bbetal}. It has been shown previously 
\cite{fehnor} that depending on the symmetry of the 
superconducting order parameter, the pair breaking rate 
or the variation of the experimentally measured $T_c$ would 
be different. This is due to the fact that a pairing 
state of reduced symmetry is unstable to impurity scattering 
to a very high degree, i.e., depression of $T_c$ is significantly 
stronger for a superconductor with $d$-wave symmetry than for 
a superconductor with pure $s$-wave symmetry.  

The issue of the symmetry of the superconducting order parameter 
(OP) in HTLS has attracted considerable attention in the recent 
past and has been discussed widely in the high-$T_c$ literature,  
a large fraction of which reported evidence of a $d_{\rm x^2-y^2}$ 
OP symmetry \cite{dwpap1,dwpap2}, and a concensus seems emerging
in this direction.
But, the striking results relating the pseudogap, as reported 
in NMR \cite{psnmr,william1,william2}, optical conductivity 
\cite{homes,psoptc}, heat capacity \cite{loram}, transport data 
\cite{pstrans} and ARPES studies \cite{psgap}, are yet to be 
understood within an unified framework. In other words, there 
have been no consensus so far regarding the origin of the 
pseudogap and the issue regarding the role played by the 
pseudogap to affect the physical characteristics of HTLS.
In this paper, we will primarily be concerned about the 
extraction of the pseudogap and its effects on the in-plane 
(CuO$_2$ plane) physical characteristics. Our basic task is 
three-fold. First, we analyse the $T_c$-suppression data 
(by impurity substitution) in Bi2212 \cite{bbetal} and Tl2212 
\cite{bb22} within the Abrikosov-Gorkov formalism 
\cite{fehnor,ageqn,maki} and investigate the suppression of 
the in-plane density of states (DOS) for variously doped samples.  
Secondly, we extract the pseudogap magnitude at different dopings 
within a ``fermi-level density of states suppression" picture 
\cite{bcetal}. And finally we study the effects of the pseudogap 
on a measurable physical characteristic of HTLS i.e. 
suceptibility. Note that, $T_c$ suppression data, which are 
obtained by substitution of only magnetic impurities (Co) at the 
Cu-sites, are considered here. An estimation of the pseudogap 
as well as the understanding of its effects on the in-plane 
physical characteristics is of importance for the purpose 
of a consistent theoretical modelling \cite{bipdas} of HTLS 
as well as for further enlightening of the important issues 
in the subject.

The paper is organised as follows. In section-II, we describe the 
procedural steps involving the extraction of the pseudogap from the 
analysis of $T_c$-suppression data. Results are presented in 
section-III together with discussions of them. Section-IV comprises 
of a brief summary of the work done and some relevant comments. 

\vspace{10pt} 
\centerline{\large{\bf II.  Data Analysis: Estimation of the Pseudogap}} 

The variation of $T_c$ as a function of increasing Co impurity 
concentration has been reported elsewhere by B. Bandyopadhyay 
{\it et al.} \cite{bbetal}. Here, we have fitted the 
$T_c$-variation data to the Abrikosov-Gorkov (AG) equation 
\cite{ageqn} valid for the weak coupling $d$-wave 
superconductors. Within the AG formalism pair breaking
rate is inversely proportional to the DOS at fermi level
denoted by $N(E_F)$. Experimental data are fitted to the AG
relation using $N(E_F)$ as a fitting parameter and its values are
thereby found out corresponding to each doping. This DOS at the fermi
level $N(E_F)$ shows a progressive depletion towards underdoping
which could imply the opening up of a gap of growing magnitude. 
Assuming that the depletion in $N(E_F)$ is caused by a pseudogap 
$E_g$ \cite{psgap}, we calculate a DOS within the ``DOS-suppression
picture" using a phenomenological form of the quasiparticle 
energy proposed previously \cite{william2,loram}. Values of the
magnitude of $E_g$ are then obtained by fitting the calculated DOS
at the fermi level (within the DOS-suppression picture) to
the $N(E_F)$ values found out 
by data analysis. The $E_g$ values thus obtained are used to 
calculate the susceptibility at different doping and experimental 
susceptibility measurement data \cite{pmandal} are compared. We 
also do a similar analysis, as above, of the $T_c$ suppression data
on Tl2212 and report the results here.

From the analysis of $T_c$ reduction data within the AG framework, 
it is found that both Bi2212 and Tl2212 cuprates bear qualitatively
similar characteristic features regarding the pseudogap and other 
related properties investigated here. It is evident in the data 
analysis that the rate of suppression of $T_c$ becomes faster towards 
underdoping as reported earlier by other groups \cite{tallon,william1}. 
We find that the magnitude of $E_g$, as obtained within 
the DOS suppression picture, grows as an inverse power of the carrier
concentration towards underdoped side and is qualitatively similar
to a related temperature scale $T^\star$ estimated by thermoelectric
power measurements \cite{bbetal}.

For Bi2212 cuprates the sample composition is given by 
Bi$_2$Sr$_2$Ca$_{1-x}$Y$_x$(Cu$_{1-y}$M$_y$)$_2$O$_8$,  
where M represents cobalt (Co) impurity and $y$ gives the 
concentration of impurity in percentage. Here $x$ is the amount 
of yttrium (Y) substituted in place of calcium (Ca) which controls 
the carrier concentration. Details of sample preparation and 
measurements of $T_c$ versus $y$ (for Bi2212) are presented 
earlier \cite{bbetal} and here we consider an analysis of the 
$T_c$ suppression data. 

For Tl-cuprate the sample composition is
Tl$_2$Ba$_2$Ca$_{1-x}$Y$_x$(Cu$_{1-y}$M$_y$)$_2$O$_8$ where
M and $y$ again represent Co impurity and its concentration 
respectively. Values of $T_c$ for both Bi2212 \cite{bbetal} and 
Tl2212 \cite{bb22} at different impurity levels are estimated 
by studying the temperature variation of the resistivity. It 
should be mentioned that the carrier concentration for the 
samples used here are changed by substitution of yttrium (Y) 
at Ca sites \cite{pmandal,mandrus}. This procedure of changing 
carrier concentration does not affect the CuO$_2$ plane
responsible for superconductivity. Whereas a change of the
carrier concentration by oxygenation procedure might directly affect 
the CuO$_2$ plane by leaving traces of inherent impurities during
sample preparation. These impurities are not accounted for while
studying the $T_c$ suppression by impurity substitution and hence,
the samples (of varied carrier concentration) prepared by Y-substitution
procedure are better suited for the purpose. Thus, $T_c$ suppression
data used here \cite{bbetal,bb22} for analysis are expected to be more 
reliable.
 
\vspace{10pt} 
\centerline{\large{\bf III.  Results and Discussions}} 

In Fig.1 we present $T_c$ versus $y$ data (symbols) for various values 
of $x$ (corresponding to different doping levels) where Fig.1(a) 
includes data for Bi2212 and Fig.1(b) gives that for Tl2212 samples. 
Clearly, for overdoped situation (small $x$) the reduction of $T_c$ 
with Co-impurity concentration ($y$) is slower than that for the
underdoped situation (large $x$). The $T_c$ reduction data are fitted
to the AG equation \cite{ageqn,maki} given as
\begin{equation} 
-\ln\left({T_c\over T_{co}}\right) 
         = \psi \left[{1\over 2} + {\Gamma\over{2\pi \kbtc}}\right] 
           - \psi \left[{1\over 2}\right], 
\end{equation} 
where $T_{co}$ is the value of $T_c$ at $y=0$, i.e., without any 
impurity scattering centre in the sample. Here $\psi[z]$ represents 
the digamma function for argument $z$ and $\Gamma = n_i/(\pi N(E_F))$
is the pair breaking scattering rate for unitary scattering. The 
quantity $N(E_F)$ is the DOS at fermi level 
(in units of /eV/$\angs^3$) and 
$n_i=\alpha y/abc$ is the density of impurity scatterers per unit 
volume where $\alpha$ is the number of CuO$_2$ planes per formula 
unit and $a$, $b$, $c$ are lattice constants. A justification for 
using the AG equation for the short coherence length cuprate 
superconductors can be found in references \cite{tallon} and 
\cite{loram}. In Fig.1(a) 
and Fig.1(b) solid and dashed lines are fitted curves using Eq.(1). 
Matching of the fitted lines with the experimental data are quite
good, as seen in Fig.1. Notice, that the experimental data points 
($T_c$ versus $y$) exist only upto a limited value of the impurity 
concentration $y=7-8\,\%$. This is because samples of a pure single 
phase are not available \cite{bbetal} beyond this value of impurity 
concentration, $y>8\%$. 

In the curve fitting process, $N(E_F)$ is taken as a fitting parameter 
and its values are found out by a least square fitting of $T_c$ versus 
$y$ data. Fitted $N(E_F)$ values corresponding to different doping 
levels (different $x$) are listed in table-I. It may be noted here 
that the typical r.m.s. deviation in the least square fitting of data 
is of the order of $1\, K$ which reaffirms the visual excellence 
of data fitting curves presented in Fig.1.  

The ratio $|$d$T_c$/d$y|$ is calculated for the fitted curves 
($T_c$ versus $y$) of Fig.1. It is seen that with the increase 
of $y$, the slope $|$d$T_c$/d$y|$ increases monotonically 
for small $y$ and then registers a steep rise before falling sharply 
to zero at some value of $y$ denoted as the critical value of the impurity 
concentration $y_c$. Values of $y_c$ are different for different doping 
levels of the samples. For overdoping $y_c$ is considerably large and 
decreases progressively but rapidly towards underdoping. Noting the 
values of $y_c$ for each curve, one could calculate the corresponding 
critical values of the pair breaking scattering rate as 
$\Gamma_c = \alpha y_c/(\pi N(E_F) a\,b\,c)$ which are listed in table-I. 
The AG relation used here (in Eq.(1)) is valid for the weak coupling 
$d$-wave superconductors for which the critical scattering rate (at which 
$T_c$ reduces to zero) is evaluated \cite{maki} to be 
$\Gamma_c = 0.88\, k_BT_{co}$. Values of $\Gamma_c$, calculated by  
this empirical relationship, are also presented in table-I which are within 
$1\,\%$ of the values obtained by data analysis. This correspondence 
might be taken as a crosscheck for the data fitting procedure and could 
also signify the validity of the AG relation used for the purpose. 

Next we calculate the DOS using the normal state quasiparticle
dispersion given as \cite{william2,loram} 
\begin{equation}
E_\vk = \left[\epsilon_\vk^2 + E_g(\vk)^2\right]^{1/2}, 
\end{equation} 
where $\eps_\vk$ represents the tight binding band energy for which 
we consider a realistic band structure obtained by a six parameter 
tight-binding fit [$t_o$,$t_1$,$t_2$,$t_3$,$t_4$,$t_5$] = 
[0.131,-0.149,0.041,-0.013,-0.014,0.013] $eV$ to the ARPES 
data on Bi2212 \cite{norman}. Here 
$E_g(\vk) = (E_g/2)\,|\eta_\vk|$ with $E_g$ being the pseudogap 
magnitude and $\eta_\vk = \cos k_xa - \cos k_ya$ ($a$ is the in-plane 
lattice constant of a square lattice) is the $d_{x^2-y^2}$ symmetry 
factor associated with the pseudogap \cite{psgap}. Appearence of $E_g$ 
in the quasiparticle dispersion effectively causes suppression of the 
DOS at the fermi level. In our analysis, we estimate $E_g$ at a fixed 
doping level by fitting its value such that the DOS at Fermi energy 
calculated by using relation (2) (within the ``DOS-suppression picture")
matches $N(E_F)$ found out previously
from the data analysis (see table-I). A plot of scaled 
$\overline{\rm E}_g$ (open squares) and $\overline{T}_{co}$ (solid 
triangles) as a function of carrier concentration ($p$), for Bi2212 
samples, is given in Fig.2. Here,
$E_g$ at different $p$ are scaled to its value at optimal doping 
$p=0.16$, that is $\overline{\rm E}_g = E_g(p)/E_g(p=0.16)$. $T_{co}$ 
values obtained in experimental measurements are also scaled  
similarly $\overline{T}_{co} = T_{co}(p)/T_{co}(p=0.16)$.
Solid line in the main figure, representing the locus of 
$\overline{T}_{co}$ data, is
a guide to the eye and the dashed line is a power law fit to
the $\overline{\rm E}_g$ data. Notice that the pseudogap magnitude 
$\overline{\rm E}_g$ shows a sharp rise towards underdoping
which is seen to vary as a power of inverse doping ($1/p$). For 
the case of Tl2212 samples, the data points (not included in the 
figure) show similar characteristic features \cite{noteTl} as 
in the case of Bi2212. It would be of relevance to  point out that 
in a recent experiment, Demsar {\it et al.} \cite{demetal} studied 
the gap-structure in YBCO single crystals employing real-time 
measurements of the quasiparticle relaxation dynamics. They found a 
$T$-independent pseudogap to remain dominant for the underdoped 
samples and even persists in the overdoped state with its magnitude 
being inversely proportional to doping. 

In the inset of Fig.2, we include the 
variation of $T^\star$ versus $p$ (solid squares) as estimated 
from TEP measurement on Bi2212 where $T^\star$ is the temperature 
at which thermoelectric power shows a peak and is much higher 
than the corresponding superconducting transition temperature 
$T_c$. The quantity $T^\star$ can be thought of as an energy scale 
related to the pseudogap $E_g$, although in some places it has 
been termed as the temperature where the pseudogap opens up 
\cite{emery}. Solid line in the inset is a fit to the data and 
depicts that $T^\star$ grows towards underdoping as a power 
of $1/p$, although the power is different from that involving 
$\overline{\rm E}_g$ versus $p$ in the main figure. This 
similarity observed within $T^\star$ and $\overline{\rm E}_g$ 
emboldens the idea that $T^\star$ is an energy scale related to 
$E_g$. It is important to note that $E_g$ values towards overdoping 
do not fall to zero, but shows monotonic decrease or remain 
nearly flat implying existence of $E_g$ even in the overdoped 
cuprates. This feature is visible in both the main figure as well
as in the experimental $T^\star$ data presented in the inset. 
Similar results of the existence of pseudogap in the overdoped 
region has been reported in a recent experiment on YBCO samples 
\cite{demetal} and for the case of Bi$_2$Sr$_2$(Ca,Y)Cu$_2$O$_8$ 
as observed in several other experiments \cite{deuts,renner}. 
This is also consistent with earlier experimental observations 
in Hg-cuprates \cite{hgexpt}. 

With the assumption, that quasiparticles are well defined in the 
vicinity of the fermi surface, susceptibility can be written 
involving DOS as 
\begin{equation} 
\chi = 2\,\mu_B^2\,\int_{-\infty}^\infty 
       \left[-\frac{\partial f(E)}{\partial E} \right] \,N(E)\,\, {\rm d} E 
\end{equation} 
where $N(E)$ is the DOS at energy $E$ and 
$f(E) = (e^{\beta E} + 1)^{-1}$ 
is the fermi distribution function. Susceptibility $\chi$ becomes 
temperature dependent through the fermi function $f(E)$. Using the 
quasiparticle dispersion as in Eq.(2), we calculate $\chi$ for 
different carrier concentrations (different $E_g$). A plot of 
$\chi/\mu_B^2$ as a function of $T$ is given in Fig.3. The 
quantity $\chi/\mu_B^2$ has the dimension of DOS denoted by $N(E)$ 
in Eq.(3). In each curve in the plot, towards low temperature 
$\chi$ falls off rapidly and tends to zero in the limit 
$T\to 0$. With increasing temperature, $\chi$ increases
slowly, peaks at a temperature ($T_\chi$) and then decreases slowly 
towards its high-$T$ saturation. Occurance of the broad peak in 
$\chi$ at high $T$ (above the corresponding experimentally measured
$T_{co}$) is due to the presence of $E_g$ in the quasiparticle 
energy and is not observable in case of a normal metal. As one 
goes towards reduced doping level, $E_g$ increases, the value of 
$T_\chi$ gets shifted towards higher temperatures and the peak 
becomes more and more broader. In the inset of Fig.3, we include 
the experimental data \cite{pmandal} of $\chi$ versus $T$ for Bi2212 
which shows the feature that $\chi$ has a broad peak above $T_c$ and 
the peak position shifts towards higher $T$ with underdoping. This 
consistency of the earlier experimental data with that calculated 
using the quasiparticle dispersion in Eq.(2) lends some support to 
the physical picture that the existence of $E_g$ affects the 
in-plane (CuO$_2$-plane) charge transport by suppressing the 
quasiparticle DOS at the fermi level.   
It may be noted here that, within a similar DOS suppression picture, 
NMR data \cite{william1,william2} as well as the normal state heat 
capacity and magnetic susceptibility data \cite{loram2} on different 
HTLS materials were analysed earlier in terms of a $d$-wave pseudogap.

\vspace{10pt} 
\centerline{\large{\bf IV.  Summary and Comments}} 

To summarize, we have considered experimental $T_c$-suppression data by 
impurity substitution in Bi2212 and Tl2212 cuprates, analysed them
within Abrikosov-Gorkov formalism of {\it pair breaking 
by impurity scattering} and estimated the pseudogap magnitude within the 
DOS suppression picture. Our analysis shows that the pseudogap energy 
varies as a power of inverse doping and the variation bears close 
resemblance to that of $T^\star$ 
found by thermoelectric power measurement on Bi2212. Presence of 
$E_g$ affects the susceptibility $\chi$ by inflicting the occurance of 
a broad peak above the corresponding experimental $T_{co}$ indicating 
an unusual metallic state. A systematic study of $\chi$ and other 
properties connected to the pseudogap, for differently doped samples 
of various layered cuprates could be useful to further explore the 
relevance of the DOS suppression (by pseudogap) viewpoint in layered 
cuprates.  

\vspace{10pt} 
\centerline{\large{\bf V.  Acknowledgements}} 
 
One of us, B. Chattopadhyay thanks CSIR, Government of India, 
for a Senior Research Associateship (Pool Scheme).

\newpage 
\baselineskip22pt
\centerline{\bf Table-I~~~~~~~} 

\begin{tabbing}
aaaaaaaaaaa\=aaaaaaa\=aaaaaaaa\=aaaaaaaaaaaa\=aaaaaaaaaaaaaaaaaaaaa\= \kill
\noindent\rule{6.05in}{0.05mm}\\ 
{Material} \> {~$x$} \> {~~$p$} \> ~{$N(E_F)$} \>\underline{~~~~~~~~~~~~~~~$\Gamma_c$ values (in meV)~~~~~~~~~~~~~~} \> \\ 
\> \> \>(/eV/$\angs^3$) \>From curve fitting \> From the relation \\ 
\> \> \> \> \> $\Gamma_c$=0.88 k$_BT_{co}$ \\ 
\noindent\rule{6.05in}{0.05mm} \\ 
Bi2212\> 0.0 \> 0.225 \> 0.0805 \> ~~~5.90 \> ~5.86 \\ 
      \> 0.2 \> 0.160 \> 0.0435 \> ~~~6.77 \> ~6.73 \\ 
      \> 0.3 \> 0.135 \> 0.0250 \> ~~~7.09 \> ~6.98 \\ 
      \> 0.4 \> 0.110 \> 0.0191 \> ~~~6.09 \> ~6.03 \\ 
\noindent\rule{6.05in}{0.05mm}\\ 
Tl2212\> 0.2 \> 0.102 \> 0.0547 \> ~~~5.94 \> ~5.89 \\ 
      \> 0.25 \> 0.085\> 0.0466 \> ~~~4.19 \> ~4.15 \\ 
\end{tabbing}
\vspace{-60pt}
\vskip0.8cm
\noindent\rule{6.05in}{0.05mm} 

\newpage
\baselineskip22pt

\noindent{\Large{\bf Figure captions:}}  

\begin{itemize}

\item[Fig.1.] Superconducting transition temperature $T_c$ is plotted 
   as a function of substituted Co impurity concentration for different 
   values of yttrium content $x$ as shown in the figure. Figure (a) 
   corresponds to Bi2212 and (b) corresponds to Tl2212 samples. In both 
   the panels, the symbols represent experimentally measured $T_c$ values 
   and lines (solid and dashes) represent fitted curves using the 
   AG equation (1) as written in the text.  
\vskip 0.3 cm

\item[Fig.2.] Phase diagram of bilayer bismuth cuprate. Solid triangles 
   are scaled $T_{co}$ values corresponding to different carrier 
   concentration ($p$) as measured experimentally. Solid line is a 
   parabolic fit to these experimental data and serves as a guide to 
   the eye. Open squares are scaled values of the pseudogap magnitude 
   ($\overline{\rm E}_g$) as estimated within the DOS suppression 
   picture. Dashed line is a power law fit to $\overline{\rm E}_g$ 
   data. [Inset: Solid squares represent values of $T^\star$, as estimated 
   by thermoelectric power measurement, as a function of $p$. 
   Solid line is a power law fit to the data.]   
\vskip 0.3 cm

\item[Fig.3.] Susceptibility for Bi-cuprate, as calculated using Eq.(3) of 
   the text, is plotted as a function of temperature for three different 
   values of $x$ mentioned in the figure. The temperature $T_P$ for each 
   curve, at which susceptibility has a broad peak, is marked by an arrow 
   and the corresponding value is written. [Inset: Susceptibility data for 
   Bi2212 as a function of temperature corresponding to three different 
   values of yttrium (Y) concentration as noted in the figure. These data 
   are taken from Ref.\cite{pmandal}.]  
 
\end{itemize} 


\begin{thebibliography}{999}
\bibitem{noteTl}{For the case of bilayer Tl-cuprates, only two 
  sets of $T_c$ suppression data are available, none of which 
  corresponds to optimal doping. Thus, values of $E_g$ obtained 
  within DOS-suppression picture could not be scaled suitably 
  and hence not presented in Fig.3. The unscaled data show a 
  rising trend towards underdoping.}
\end{thebibliography}
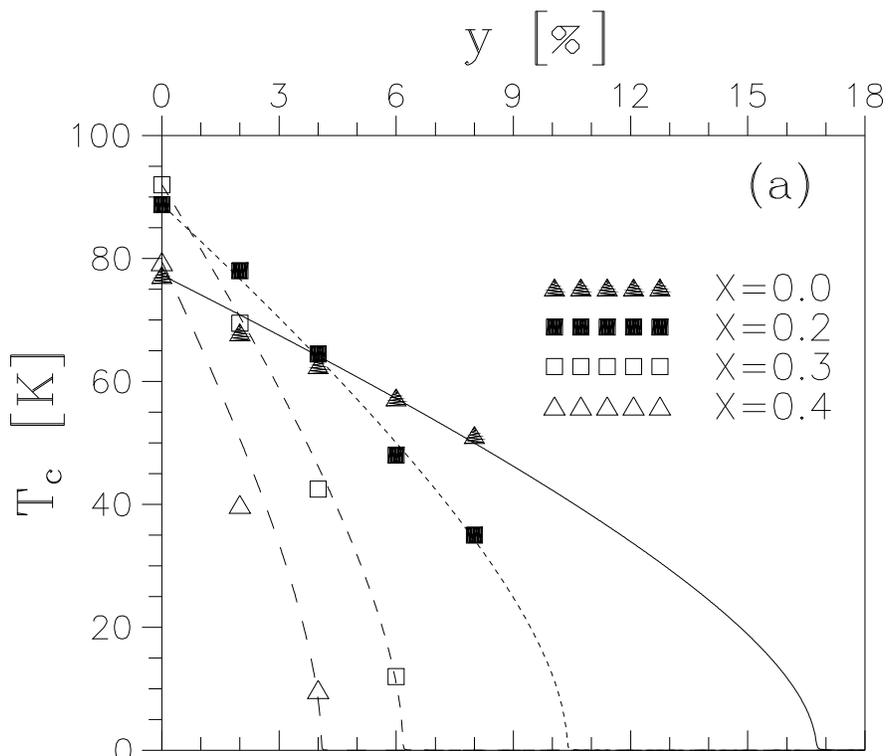
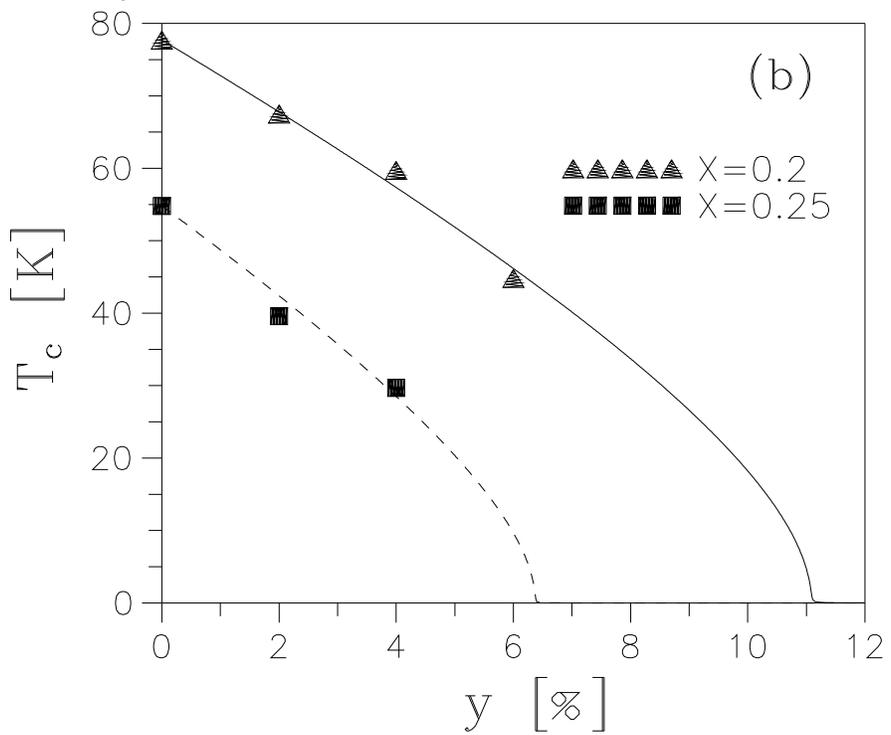
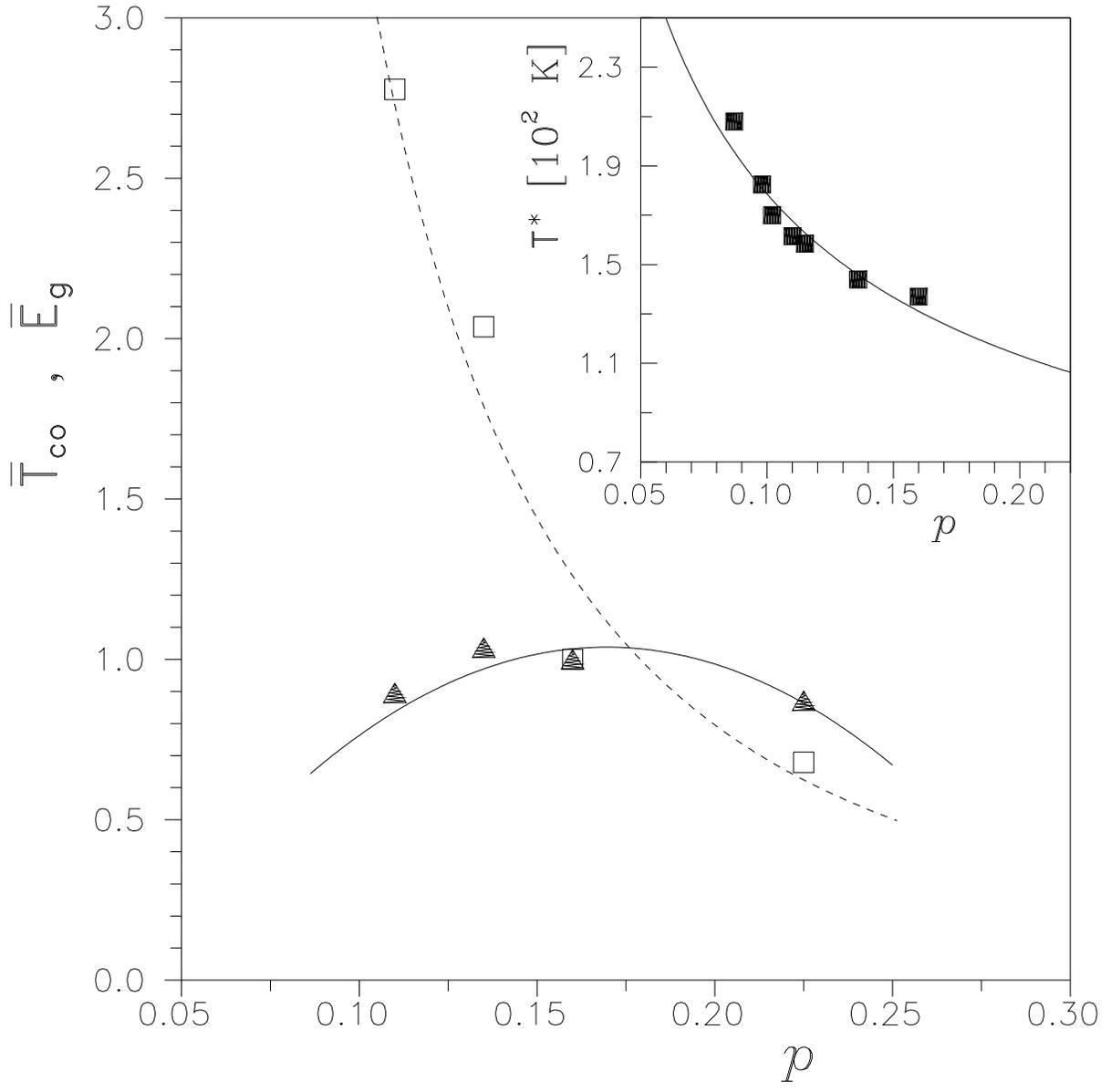
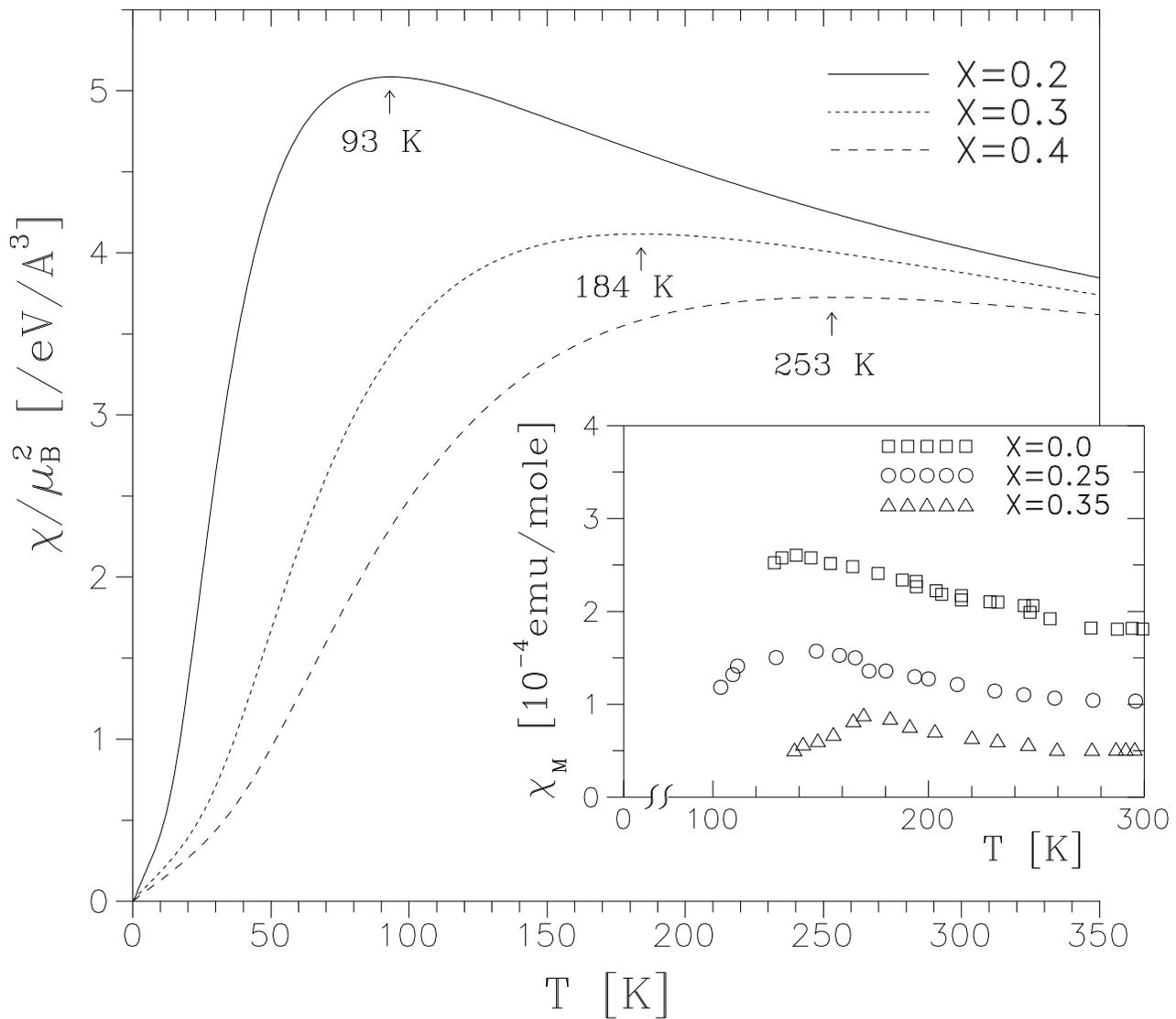
\end{document}